\documentclass[
aps,prd,
12pt,
nopreprintnumbers,
showpacs,
eqsecnum,
nofootinbib
]{revtex4-1}

\usepackage{graphicx}

\def\tr{{\rm tr\,}}
\def\vx{{\bf x}}

\def\v0{{\bf 0}}

\begin{document}

\title{
Dirac-Born-Infeld-Einstein theory with Weyl invariance
}
\author{Takuya Maki}\email[]{maki@jwcpe.ac.jp}
\affiliation{
Japan Women's College of Physical Education, 
Setagaya,~Tokyo 157-8565,~Japan}
\author{Nahomi Kan}\email[]{kan@yamaguchi-jc.ac.jp}
\affiliation{
Yamaguchi Junior College,
Hofu-shi, Yamaguchi 747--1232, Japan}
\author{Kiyoshi Shiraishi}\email[]{shiraish@yamaguchi-u.ac.jp}
\affiliation{
Yamaguchi University,
Yamaguchi-shi, Yamaguchi 753--8512, Japan}
\date{\today}

\begin{abstract}
Weyl invariant gravity has been investigated as the fundamental theory of
the vector inflation. Accordingly, we consider a Weyl invariant extension
of Dirac-Born-Infeld type gravity. We find that an appropriate choice of
the metric removes the scalar degree of freedom which is  at the first
sight required by the local scale invariance of the action, and then a
vector field acquires mass.  Then nonminimal couplings of the vector
field and curvatures are induced. We find that the Dirac-Born-Infeld type
gravity is a suitable theory to the vector inflation scenario.
\end{abstract}



\maketitle

\section{Introduction}

The cosmological inflation is proposed as some resolutions for the
important cosmological problems, {\it e.g.} the flatness, horizon and
monopole problems. Most of successful models are based on
classical scalar fields, although
we have not observed such scalar bosons associated with the field. 

The inflation can also be caused by other type of fields. The
vector inflation has been proposed  by Ford~\cite{F} and some
authors~\cite{BL,GMV,GV}. 
It is emphasized that the massive vector field should
non-minimally couple to gravity in
such models~\cite{F,BL,GMV,GV}.

The reason why the nonminimal coupling is important is as follows.
Suppose the equation of motion for the vector field is given by
\begin{equation}
\frac{1}{\sqrt{-g}}\partial_\mu(\sqrt{-g}F^{\mu\nu})-
\left(m^2-\frac{R}{6}\right)A^\nu=0\,.
\label{eom}
\end{equation}
For the background field, we assume%
\footnote{Of course, only the vector field as the source cannot lead to
the exactly isotropic expansion.}
\begin{equation}
ds^2=-dt^2+a^2(t)d\vx^2\,,
\end{equation}
and
$A_i~~(i=1, 2, 3)$ depends only on $t$ and $A_0\equiv 0$.
Then we define $B_i\equiv \frac{1}{a}A_i$, (\ref{eom}) becomes%
\begin{equation}
\ddot{B}_i+3\frac{\dot{a}}{a}\dot{B}_i
+m^2
B_i=0\,,
\label{1.3}
\end{equation}
which is very similar to the equation for a homogeneous scalar field
in the Friedmann-Lema\^{\i}tre-Robertson-Walker universe. Moreover, the
energy density is expressed as $\sim \dot{B}_i^2+m^2B_i^2$, which is also
similar to the one for the scalar field. Thus the slow evolution of
the effective scalar field $B_i$ can occur in the approximately isotropic
inflating universe.

We have studied~\cite{MNS} Weyl invariant
gravity~\cite{Weyl,Deser,SD,D,PGOF,Utiyama,Kasuya,Kugo,R,Cheng,Kao,HP,WP,P,NR,WC,JM,Scholz}
as a candidate for the theoretical model of the vector inflation.  
We found that the choice of the frame yields the mass of the Weyl gauge
field, but the nonminimal coupling term is lost~\cite{MNS}.
We come to the conclusion that we need further generalization of the
gravitational§ theory.

In the different context, Deser and Gibbons
considered Dirac-Born-Infeld (DBI)-Einstein theory~\cite{DG}  almost a
decade ago, whose Lagrangian density is of the following type
\begin{equation}
\pm\sqrt{-\det(g_{\mu\nu}\pm\alpha R_{\mu\nu})}\,,
\label{origDG}
\end{equation}
where $R_{\mu\nu}$ is the Ricci tensor and the $\alpha$ is a constant.
Originally, electromagnetism of the DBI type has been
considered as a candidate of the nonsingular theory of electric fields.
Therefore the Dirac-Born-Infeld-Einstein
theory as the highly-nonlinear theory is also expected as a theory of
gravity suffered from no argument of singularity.
The studies on the theory have been done by many
authors~\cite{W,Vo,Nieto,DC2,DC1,Rojas,GST}.
Because of the nonlinearity in this theory, we expect the extension as
the theory of gravity which realize a successful vector inflation.

Consider the Weyl invariant $D$-dimensional extension of the Ricci
curvature  (see the next section) is
\begin{equation}
\tilde{R}_{\nu\sigma}[g,A]\equiv R_{\nu\sigma}+
F_{\nu\sigma}-\left[
(D-2)\nabla_\sigma A_\nu+
g_{\nu\sigma}\nabla_\mu A^\mu
\right]+(D-2)\left(
A_\nu A_\sigma
-A_\lambda A^\lambda
g_{\nu\sigma}\right)
\,.
\end{equation}
If simple replacement of the Ricci tensor by the Weyl invariant tensor in
the action (\ref{origDG}), the expansion
\begin{equation}
\sqrt{\det{(1+A)}}
=1+\frac{1}{2}\tr A
+\frac{1}{8}((\tr A)^2-2\tr A^2)
+\cdots\,,
\label{expa}
\end{equation}
yields the terms $RA_\mu A^\mu$ and $R_{\mu\nu}A^\mu A^\nu$ and so on as
well as $R$ and $F_{\mu\nu}F^{\mu\nu}$.
Other Weyl invariant terms are necessary,
because the metric tensor must be combined with a scalar field which
compensates the dimensionality. After the frame choice, the freedom of
the scalar field is eaten by the vector field, then the presence of the
nonminimal terms mentioned above is still realized.%
\footnote{Note that $f(\tilde{R}[g,A])$ does not bring about substantial
nonminimal couplings~\cite{MNS}.}

In the next section, we review the Weyl invariant
gravity with the vector field~\cite{Utiyama,Kasuya,Kugo,Cheng,Kao,NR,WC,JM}. 
The expression (\ref{origDG}) is generalized to the Weyl invariant one. 
The Lagrangian for a Weyl-invariant DBI gravity is proposed in
Sec.~\ref{sec3}. In
Sec.~\ref{sec4}, the necessity condition for the vector inflation is
investigated. In Sec.~\ref{sec5}, another possible inflationary scenario
is provided. The last section is devoted to the summary and prospects.

\section{Weyl's gauge gravity theory
\label{sec2}}

In this section, we review the Weyl's gauge transformation
to construct the gauge invariant Lagrangian.

Consider the transformation of metric (in $D$ dimensions)
\begin{equation}
g_{\mu\nu}\rightarrow g'_{\mu\nu}=e^{2\Lambda(x)}g_{\mu\nu}\,,
\label{trans_metric1}
\end{equation}
where $\Lambda(x)$ is an arbitrary function of the coordinates $x^\mu$.

We can define the field with weight $d=-\frac{D-2}{2}$ which transforms as
\begin{equation}
\Phi\rightarrow \Phi'=e^{-\frac{D-2}{2}\Lambda(x)}\Phi\,.
\end{equation}

In order to construct the locally invariant theory, we consider the
covariant derivative of the scalar field 
\begin{equation}
\tilde\partial_\mu\Phi\equiv
\partial_\mu\Phi-\frac{D-2}{2}\,A_\mu\Phi\,,
\end{equation}
where $A_\mu$ is a Weyl's gauge invariant vector field.

Under the Weyl gauge field transformation
\begin{equation}
A_\mu\rightarrow A'_\mu=
A_\mu-\partial_\mu\Lambda(x)\,,
\label{gaugetrans}
\end{equation}
we obtain the transformation of the covariant derivative of the 
scalar field as
\begin{equation}
\tilde\partial_\mu\Phi\rightarrow
e^{-\frac{D-2}{2}\Lambda(x)}\tilde\partial_\mu\Phi\,.
\end{equation}

The field
strength of the vector field is given by
\begin{equation}
F_{\mu\nu}\equiv\partial_\mu A_\nu-\partial_\nu A_\mu\,,
\end{equation}
which is gauge invariant as
\begin{equation}
F_{\mu\nu}\rightarrow F'_{\mu\nu}=F_{\mu\nu}\,.
\end{equation}
The modified Christoffel symbol is defined as
\begin{equation}
\tilde\Gamma^\lambda_{\mu\nu}\equiv\frac{1}{2}g^{\lambda\sigma}
\left(\tilde\partial_\mu
g_{\sigma\nu}+\tilde\partial_\nu
g_{\mu\sigma}-\tilde\partial_\sigma
g_{\mu\nu}\right)\,,
\end{equation}
where $\tilde{\partial}_\mu g_{\sigma\mu}\equiv{\partial}_\mu
g_{\sigma\mu}+2A_\mu g_{\sigma\mu}$.
 The modified curvature is given as follows:
\begin{equation}
\tilde{R}^\mu{}_{\nu\rho\sigma}[g, A]\equiv
\partial_\rho\tilde\Gamma^\mu_{\nu\sigma}-
\partial_\sigma\tilde\Gamma^\mu_{\nu\rho}+
\tilde\Gamma^\mu_{\lambda\rho}\tilde\Gamma^\lambda_{\nu\sigma}-
\tilde\Gamma^\mu_{\lambda\sigma}\tilde\Gamma^\lambda_{\nu\rho}\,.
\end{equation}
The Ricci curvature in the Weyl invariant version is
\begin{eqnarray}
&
&\tilde{R}_{\nu\sigma}[g,A]\equiv\tilde{R}^\mu{}_{\nu\mu\sigma}[g,A]
=R_{\nu\sigma}+
F_{\nu\sigma}+\left(
\nabla_\sigma A_\nu-
D\nabla_\sigma A_\nu-
g_{\nu\sigma}\nabla_\mu A^\mu+
\nabla_\sigma A_\nu
\right)\nonumber \\
& &\qquad\qquad\qquad+\left[
A^\mu\left(g_{\nu\sigma}A_\mu-g_{\nu\mu}A_\sigma\right)-
A_\nu\left(A_\sigma-DA_\sigma\right)
-A_\lambda A^\lambda \left(D
g_{\nu\sigma}-
g_{\nu\sigma}\right)\right]\nonumber \\
& &\qquad =R_{\nu\sigma}+
F_{\nu\sigma}-\left[
(D-2)\nabla_\sigma A_\nu+
g_{\nu\sigma}\nabla_\mu A^\mu
\right]+(D-2)\left(
A_\nu A_\sigma
-A_\lambda A^\lambda
g_{\nu\sigma}\right)
\,,
\end{eqnarray}
where $\nabla$ denotes the usual generally covariant derivative.
Note that under the gauge transformation
\begin{equation}
\tilde{R}_{\nu\sigma}[g,A]\rightarrow\tilde{R}_{\nu\sigma}[g',A']=
\tilde{R}_{\nu\sigma}[g,A]\,.
\end{equation}

\section{Weyl invariant Lagrangian
\label{sec3}}

Although we can use the Weyl invariant Ricci tensor $\tilde{R}_{\mu\nu}$
in the DBI gravity,
we should note that the metric tensor in the action is not Weyl invariant
(which is shown in (\ref{trans_metric1})). Thus, we use a combination 
$\Phi^\frac{4}{D-2}g_{\mu\nu}$
 instead of the
metric tensor. The scalar $\Phi$ compensates the dimension of the metric. 
Now the use of $\tilde{R}_{\mu\nu}$ and  $\Phi^\frac{4}{D-2}g_{\mu\nu}$ 
in the DBI type action leads to the theory of gravity, a vector field,
and unexpectedly, a scalar field.

The introduction of the compensating scalar field 
tell us the action is far from general one.
The monomial of the type of the kinetic term, in other words, 
two coordinate derivatives of the scalar field can be considered, while
the curvature includes also two derivatives with no contraction.
The possible monomials are
\begin{equation}
\Phi^{-2}\tilde{\partial}_\mu\Phi\tilde{\partial}_\nu\Phi\qquad {\rm
and}\qquad
\tilde{\nabla}_\mu(\Phi^{-1}\tilde{\partial}_\nu\Phi)
\,.
\end{equation}

Another notice is in order.
The decomposition of a rank two tensor show that
there are three irreducible ones;
an antisymmetric tensor, 
a traceless symmetric tensor and a trace part. 

Now we must introduce the following independently Weyl invariant tensors
into the determinant in the DBI theory:
\begin{eqnarray}
& &\Phi^\frac{4}{D-2}g_{\nu\sigma}\,,\quad
\tilde{R}^S_{\nu\sigma}[g,A]\,,\quad
\tilde{R}[g,A]g_{\nu\sigma}\,,\quad
F_{\nu\sigma}\,,\quad
\Phi^{-2}\tilde{\partial}_\nu\Phi\tilde{\partial}_\sigma\Phi\,,\nonumber
\\ & &\Phi^{-2}g^{\lambda\mu}\tilde{\partial}_\lambda\Phi
\tilde{\partial}_\mu\Phi g_{\nu\sigma}\,,\quad
\tilde{\nabla}_\sigma(\Phi^{-1}\tilde{\partial}_\nu\Phi)
+\tilde{\nabla}_\nu(\Phi^{-1}\tilde{\partial}_\sigma\Phi)
\,,\quad
\tilde{\nabla}^\mu(\Phi^{-1}\tilde{\partial}_\mu\Phi)
g_{\nu\sigma}\,,
\end{eqnarray}
where
\begin{eqnarray}
& &\tilde{R}^S_{\nu\sigma}[g,A]\nonumber \\
& &=R_{\nu\sigma}-\left[
\frac{D-2}{2}(\nabla_\sigma A_\nu+\nabla_\nu A_\sigma)+
g_{\nu\sigma}\nabla_\mu A^\mu
\right]+(D-2)\left(
A_\nu A_\sigma
-A_\lambda A^\lambda
g_{\nu\sigma}\right)
\,,
\end{eqnarray}
and \begin{equation}
\tilde{R}[g,A]\equiv
g^{\nu\sigma}\tilde{R}_{\nu\sigma}[g,A]=R-2(D-1)\nabla_\mu
A^\mu-(D-1)(D-2)A_\mu A^\mu\,.
\label{R}
\end{equation}
We choose those as symmetric tensors are not traceless.%
\footnote{Judging from the number of fields and derivatives,
the term $\Phi^{-\frac{4}{D-2}}g^{\lambda\mu}F_{\nu\lambda}F_{\sigma\mu}$
is allowed in the same order. But this term is different from others in
the point that it includes two kinds of fields except for the metric.
Therefore we discarded this marginally possible term here.}

Our model of Weyl invariant DBI gravity is described by the Lagrangian
density
\begin{equation}
{\cal L}=-\sqrt{-\det M_{\mu\nu}}+(1-\lambda)\sqrt{-\det
(\Phi^{\frac{4}{D-2}}g_{\mu\nu})}\,,
\end{equation}
with
\begin{eqnarray}
M_{\mu\nu}&\equiv&\Phi^{\frac{4}{D-2}}g_{\mu\nu}-\alpha_1
\tilde{R}^S_{\mu\nu}[g,A]-\alpha_2
\tilde{R}[g,A]g_{\mu\nu}+\beta F_{\mu\nu}\nonumber \\& &
+
\gamma_1\Phi^{-2}\tilde{\partial}_\mu\Phi\tilde{\partial}_\nu\Phi
+\gamma_2\Phi^{-2}g^{\lambda\sigma}\tilde{\partial}_\lambda\Phi
\tilde{\partial}_\sigma\Phi g_{\mu\nu}\nonumber \\
& &-\gamma_3\left[\tilde{\nabla}_\mu(\Phi^{-1}\tilde{\partial}_\nu\Phi)
+\tilde{\nabla}_\nu(\Phi^{-1}\tilde{\partial}_\mu\Phi)\right]
-\gamma_4\, g^{\lambda\sigma}
\tilde{\nabla}_\lambda(\Phi^{-1}\tilde{\partial}_\sigma\Phi)
g_{\mu\nu}\,,
\end{eqnarray}
where $\alpha_1$, $\alpha_2$, $\beta$, $\gamma_1$, $\gamma_2$,
$\gamma_3$, $\gamma_4$ and $\lambda$ are dimensionless constants.%
\footnote{If we demand that the terms with lowest derivatives in the
expansion (\ref{expa}) look like the Lagrangian of scalar-tensor theory,
we must choose as
$\alpha_1+4\alpha_2>0$ and $\gamma_1+4\gamma_2+4\gamma_3+8\gamma_4>0$, for
$D=4$.}

Furthermore the Lagrangian density can be expressed by the
new metric conformally related to the original one and new variables.
Here we choose
\begin{equation}
\hat{g}_{\mu\nu}\equiv f^{-2}\Phi^{\frac{4}{D-2}}g_{\mu\nu}\,,
\end{equation}
and 
\begin{equation}
\hat{A}_\mu\equiv A_\mu-\frac{2}{D-2}\partial_\mu\ln\Phi\,.
\end{equation}
Note that a mass scale $f$ was introduced here.
By using the new metric and vector field, we rewrite the each term in the
determinant of the Lagrangian as
\begin{eqnarray}
& &\Phi^\frac{4}{D-2}g_{\nu\sigma}=f^2\hat{g}_{\mu\nu}\,,\quad
\tilde{R}^S_{\nu\sigma}[g,A]=\tilde{R}^S_{\nu\sigma}[\hat{g},\hat{A}]
\,,\quad
\tilde{R}[g,A]g_{\nu\sigma}=\tilde{R}[\hat{g},\hat{A}]\hat{g}_{\nu\sigma}
\,,\quad
F_{\nu\sigma}=\hat{F}_{\nu\sigma}\,,\nonumber \\
&
&\Phi^{-2}\tilde{\partial}_\nu\Phi\tilde{\partial}_\sigma\Phi=
\left(\frac{D-2}{2}\right)^2 \hat{A}_\nu\hat{A}_\sigma\,,\quad
\Phi^{-2}g^{\lambda\mu}\tilde{\partial}_\lambda\Phi
\tilde{\partial}_\mu\Phi g_{\nu\sigma}=\left(\frac{D-2}{2}\right)^2
\hat{g}^{\lambda\mu}\hat{A}_\lambda\hat{A}_\mu\hat{g}_{\nu\sigma}
\,,\nonumber \\
& &
\tilde{\nabla}_\sigma(\Phi^{-1}\tilde{\partial}_\nu\Phi)
+\tilde{\nabla}_\nu(\Phi^{-1}\tilde{\partial}_\sigma\Phi)=
(D-2)\left[-\frac{1}{2}(\hat{\nabla}_\nu \hat{A}_\sigma+
\hat{\nabla}_\sigma \hat{A}_\nu)+2\hat{A}_\nu\hat{A}_\sigma-
\hat{g}^{\lambda\mu}\hat{A}_\lambda\hat{A}_\mu\hat{g}_{\nu\sigma}\right]
\,,\nonumber \\
& &
g^{\lambda\mu}\tilde{\nabla}_\lambda(\Phi^{-1}\tilde{\partial}_\mu\Phi)
g_{\nu\sigma}=
(D-2)\left[-\hat{\nabla}^\mu \hat{A}_\mu-(D-2)
\hat{g}^{\lambda\mu}\hat{A}_\lambda\hat{A}_\mu\right]
\hat{g}_{\nu\sigma}\,.
\end{eqnarray}
We now can write $M_{\mu\nu}$ as
\begin{eqnarray}
M_{\mu\nu}&=&f^{2}g_{\mu\nu}-\alpha_1
{R}_{\mu\nu}-\alpha_2
{R}g_{\mu\nu}+\beta F_{\mu\nu}\nonumber \\& &
+
\gamma'_1A_\mu A_\nu
+\gamma'_2 g^{\rho\sigma}A_\rho
A_\sigma g_{\mu\nu}\nonumber \\
& &+\gamma'_3\left({\nabla}_\mu A_\nu
+{\nabla}_\nu A_\mu\right)
+\gamma'_4\, 
{\nabla}^\rho A_\rho\,
g_{\mu\nu}\,,
\end{eqnarray}
where the `hat's are dropped and dimensionless constants are
\begin{eqnarray}
\gamma'_1&=&-(D-2)\alpha_1+\left(\frac{D-2}{2}\right)^2\gamma_1-2(D-2)\gamma_3
\,,\nonumber
\\
\gamma'_2&=&(D-2)\alpha_1+(D-1)(D-2)\alpha_2+\left(\frac{D-2}{2}\right)^2\gamma_2+(D-2)\gamma_3
+(D-2)^2\gamma_4
\,,\nonumber
\\
\gamma'_3&=&\frac{D-2}{2}(\alpha_1+\gamma_3) \,,\nonumber \\
\gamma'_4&=&\alpha_1+2(D-2)\alpha_2+(D-2)\gamma_4 \,.\nonumber 
\end{eqnarray}

We can rewrite the Lagrangian as
\begin{eqnarray}
{\cal L}&=&-\sqrt{-\det M_{\mu\nu}}+(1-\lambda)f^D\sqrt{-g}
\nonumber \\
&=&-\sqrt{-{g}}\sqrt{\det M^{\mu}{}_{\nu}}+(1-\lambda)f^D\sqrt{-g}\,.
\end{eqnarray}
This is the candidate Lagrangian for the vector inflation.

\section{Cosmology of Weyl's gauge Gravity
\label{sec4}}
In this section, we apply our Weyl invariant DBI theory of gravity
to cosmology in four dimensions
($D=4$).

We take the metric for the homogeneous flat universe as
\begin{equation}
ds^2=-dt^2+a_1^2(t)dx^2+a_2^2(t)dy^2+a_3^2(t)dz^2
\end{equation}
and, moreover, we assume the approximate isotropy $a_1\approx a_2\approx
a_3=a(t)$.

We consider that only $A_1(t)$ is homogeneously evolving,
and $A_2=A_3=A_0=0$.

By these ansatze, we look for the condition that the vector field
behaves much like a scalar field at classical homogeneous level.
Substituting the ans\"atze, we find
\begin{eqnarray}
M^{0}{}_{0}&=&f^2-3\alpha_1\frac{\ddot{a}}{a}-6\alpha_2
\left[\frac{\ddot{a}}{a}+\left(\frac{\dot{a}}{a}\right)^2\right]
+\gamma'_2\frac{1}{a^2}A_1^2\,,
\\
M^{0}{}_{1}&=&-\beta\dot{A}_1-\gamma'_3\left(
\dot{A}_1-2\frac{\dot{a}}{a}A_1\right)\,,\\
M^{1}{}_{0}&=&-\beta\frac{\dot{A}_1}{a^2}+\gamma'_3\frac{1}{a^2}\left(
\dot{A}_1-2\frac{\dot{a}}{a}A_1\right)\,,\\
M^{1}{}_{1}&=&f^2-\alpha_1\left[\frac{\ddot{a}}{a}+2\left(\frac{\dot{a}}{a}\right)^2
\right]-6\alpha_2
\left[\frac{\ddot{a}}{a}+\left(\frac{\dot{a}}{a}\right)^2\right]
+(\gamma'_1+\gamma'_2)\frac{1}{a^2}A_1^2\,,\\
M^{2}{}_{2}&=&M^{3}{}_{3}=f^2-\alpha_1\left[\frac{\ddot{a}}{a}+2\left(\frac{\dot{a}}{a}\right)^2
\right]-6\alpha_2
\left[\frac{\ddot{a}}{a}+\left(\frac{\dot{a}}{a}\right)^2\right]
+\gamma'_2\frac{1}{a^2}A_1^2\,.
\end{eqnarray}

After some calculations, we can subtract the part of the Lagrangian which
includes bilinear and higher-order of the vector field $A_1$.
We find that if the parameters are chosen as
\begin{equation}
\beta^2=\frac{1}{2}\left(
5\alpha_1\gamma'_1+12\alpha_2\gamma'_1+12\alpha_1\gamma'_2+48
\alpha_2\gamma'_2\right)\,,
\label{co1}
\end{equation}
and
\begin{equation}
(\gamma'_3)^2=-\frac{1}{2}\alpha_1\gamma'_1\,,
\label{co2}
\end{equation}
the vector-field part becomes 
\begin{equation}
a^3\left[\frac{1}{2}(\beta^2-\gamma_3'^2)\dot{B}_1^2-
\frac{f^2}{2}(\gamma_1'+4\gamma_2')B_1^2-
\frac{1}{8}\left(-\gamma_1'^2+4\gamma_1'\gamma_2'+8\gamma_2'^2\right)
B_1^4+\cdots
\right]\,,
\end{equation}
where $B_1=\frac{A_1}{a}$.

A simple case is realized when
$\alpha_2=\gamma_1=\gamma_2=\gamma_3=\gamma_4=0$, or these parameter take
small values in comparison with $\alpha_1$.
Then the parameter is $\alpha_1$ only.
Equations (\ref{co1}) and (\ref{co2}) tell us
$\gamma'_1=-\gamma'_2=-2\alpha_1$, $\gamma'_3=\alpha_1$ and
$\beta^2=7\alpha_1^2$. In this case, this is so simple that the effective
mass for
$B_1$ may be large. The tuning is possible; say, the choice of $\gamma_4$
does  not affects (\ref{co2}) and makes the change in the effective mass.

An elaborate tuning may give the potential which induces the chaotic
inflation~\cite{chaotic}.
In the next section, however, we show another simple inflation scenario.

\section{a simple cosmological scenario
\label{sec5}}

The chaotic inflation in the model can occur by tuning of the parameters.
We should remember that the model involves the higher-derivative
gravity. Therefore another kind of inflation is worth to be considered.

First let us suppose the flat space.
Then the potential, or the energy density for the constant $B_1$, can be
easily written down as
\begin{equation}
V=\sqrt{(f^2+\gamma'_2B_1^2)^3(f^2+(\gamma'_1+\gamma'_2)B_1^2)}\,.
\end{equation}
Although other choices are possible, we consider here a simple choice as
$\gamma'_1=0$ and $\gamma'_2<0$.%
\footnote{Note that $\gamma'_2$ can be tuned by take an
appropriate value for $\gamma_4$.}
In this case, unfortunately, the previous conditions
(\ref{co1},\ref{co2}) cannot be satisfied simultaneously, because
$\alpha_1+4\alpha_2>0$ for the positive coefficient of the
Einstein-Hilbert term in the action. 
Then the potential is
\begin{equation}
V={(f^2-|\gamma'_2|B_1^2)^2}\,.
\end{equation}
This is the simplest potential. In the true vacuum, the vector field
`condensates' and a `natural' choice $\lambda=1$ leads to vanishing
cosmological constants!%
\footnote{The parameters $\gamma'$s can be taken to be sufficiently small
so that no `antigravity' emerges.}

This simplest version also has an inflationary phase.
That is, for $B_1=0$, the scale factor behaves as $a(t)\approx e^{Ht}$
where $H^2=f^2/(3(\alpha_1+4\alpha_2))$.

Unfortunately, this phase is stabilized by the nonminimal coupling
between curvatures and the vector field.
\begin{equation}
V=\sqrt{|\gamma'_2|^2B_1^4(|\gamma'_2|^2B_1^4+4(\gamma'_3)^2H^2B_1^2)}\,.
\end{equation}
The exit of the de Sitter phase is problematic, like the other
higher-derivative models. Though the additional matter fields may play
important roles, we will perform further study on them elsewhere.

\section{Summary and outlook
\label{summary}}
The Weyl invariant DBI gravity is a candidate for a model which causes an
inflationary universe.  If the vector inflation can explain the possible
anisotropy in the early universe, we may seriously investigate the Weyl
invariant DBI gravity.

Here we examined slow development of the
massive vector field.
The inflation along with a fast evolution is shown to be possible in the
DBI inflation, where the scalar degrees of freedom which originates from
string (field) theory or D brane theory~\cite{BI}. The similar scenario is
feasible in our model, though the higher-derivatives make the detailed
analysis difficult. Anyway, numerical calculations and large simulations
will be needed to understand the minute meaning of the Weyl
invariant DBI gravity, because  the local inhomogenuity in the
spatial directions as well as the strength of vector fields is important
for thorough understanding in the early cosmology.%

Finally, we think that some marginally related subjects are in order.
The higher-dimensional cosmology in the Weyl invariant DBI gravity is
worth studying because of its rich content. %
Incidentally, DBI gravity in three dimensions is eagerly
studied~\cite{tekin}, which is related to New Massive Gravity~\cite{NMG}.
We think that the Weyl invariant extension of the lower-dimensional
theory is also of much mathematical interest.

\section*{Note added}
After completing this manuscript, we become aware of the paper ``Higgs mechanism for New Massive
gravity and Weyl invariant extensions of higher derivative theories''
by Dengiz and Tekin \cite{ntekin}.
They investigated a Weyl-invariant DBI gravity in three dimensions.

We also become aware of two recent papers about the cosmology of Weyl invariant theory~\cite{moon}.

\begin{acknowledgments}
The authors would like to thank the organizers of JGRG20, where our
partial result ({\tt [arXiv:1012.5375]}) was presented.
We also thank  D.~Comelli, S.~Deser, T.~Moon and B.~Tekin for informing us about their important
and interesting papers.

The present study is supported in part by the Grant-in-Aid of Nikaido
Research  Fund.
\end{acknowledgments}



\bibliographystyle{apsrev4-1}

\end{document}